\documentclass[a4paper,fleqn]{cas-dc}
\usepackage{longtable}
\begin{document}
\let\WriteBookmarks\relax
\def\floatpagepagefraction{1}
\def\textpagefraction{.001}

\shorttitle{Nonlinear scintillation effects in the intrinsic luminescence from Sc$_{1.318}$Y$_{0.655}$Si$_{1.013}$O$_{4.987}$ crystal excited by electrons and $\gamma$-quanta }    

\shortauthors{M. V. Belov, V. A. Kozlov, N. V. Pestovskii et al}  

\title [mode = title]{Nonlinear scintillation effects in the intrinsic luminescence from Sc$_{1.318}$Y$_{0.655}$Si$_{1.013}$O$_{4.987}$ crystal excited by electrons and $\gamma$-quanta}  

\author[1]{M.~V.~Belov}
\ead{belovmv@lebedev.ru} 
\affiliation[1]{organization={P.N. Lebedev Physical Institute of the Russian Academy of Sciences},
            addressline={Leninskii prospekt 53}, 
            postcode={119991}, 
            state={Moscow},
            country={Russia}}
            
\author[1]{V.~A.~Kozlov}
\ead{kozlov@lebedev.ru}

\author[1]{N.~V.~Pestovskii}
\cormark[1]
\cortext[1]{Corresponding author}
\ead{pestovskii@lebedev.ru}
\ead{pestovsky@phystech.edu}

\author[1]{S.~Yu.~Savinov}
\ead{savinov@lebedev.ru}

\author[1]{V.~S.~Tskhay}
\ead{vtskhay@lebedev.ru}

\author[2]{V.~I.~Vlasov}
\ead{vlasov@lsk.gpi.ru}
\affiliation[2]{organization={A.M. Prokhorov General Physics Institute of the Russian Academy of Sciences},
            addressline={Vavilova str. 38}, 
            postcode={119991}, 
            state={Moscow},
            country={Russia}}

\author[2]{A.~I.~Zagumennyi}
\ead{zagumen@lsk.gpi.ru}

\author[2]{Yu.~D.~Zavartsev}
\ead{zavart@lsk.gpi.ru}   

\author[1]{M.~V.~Zavertyaev}
\ead{zavert@lebedev.ru }
            


\begin{abstract}
The spectral and kinetic properties of intrinsic luminescence from (Y$_{2}$Sc$_{1}$)$_{0.(3)}$(Sc)[Si]O$_{5}$ crystal are studied. The emission is excited by electrons and $\gamma$-quanta. The composition (Y$_{2}$Sc$_{1}$)$_{0.(3)}$(Sc)[Si]O$_{5}$ is the congruent one for Sc$_{2}$SiO$_{5}$-Y$_{2}$SiO$_{5}$ solid solutions. It is found, that the crystal emits fairly bright intrinsic cathodololuminescence (CL) and radioluminescence (RL) at room temperature. In particular, the light yield of scintillation excited by $\gamma$-quanta with the energies of 661.7 keV is of 12000 photons/MeV. An increase in the beam flux by $\sim$20 times leads to the shift in the maximum CL energy spectral density from 315 to 340 nm and to the decrease in the CL decay time at 415 nm from 1377 $\pm$ 3 ns to 1165 $\pm$ 1 ns. Simultaneously, the decay time of RL excited by a photoelectron with the energy of 644.7 keV is of 1310 $\pm$ 10 ns while a Compton electron with the energy of 477 keV excites RL with the decay time of 1050 $\pm$ 10 ns. Also, we observed differences in the CL yield dependencies on the volume-averaged density of electronic excitations (EEs) at different wavelengths. An explanation of the results is given considering the nonlinear scintillation phenomena induced by an interaction between EEs. It is based on a conception that an increase in EE volume density leads to an increase in EEs nonradiative quenching due to these interactions. 
\end{abstract}


\begin{highlights}
\item Bright intrinsic luminescence from a congruent (Y$_{2}$Sc$_{1}$)$_{0.(3)}$(Sc)[Si]O$_{5}$ compound is found
\item Nonlinear effects in both cathodoluminescence and radioluminescence are analyzed
\item A shift in the emission spectrum and a change in the decay time are observed
\item The excitation densities are estimated and their effect on scintillation is discussed

\end{highlights}


\maketitle

\section{Introduction}\label{Introduction}

	Ionizing radiation produces in a matter the ionization tracks – spatial areas where the most of short-lived electronic excitations (EEs) are concentrated. Most of them are non-equilibrium electrons and holes. A radiative decay of these states leads to scintillation – luminescence from media excited by ionizing radiation. Due to a separation of electrons and holes and their subsequent recombination, scintillation is inherently nonlinear process \cite{1}. Indeed, in general, the scintillation energy of a unit volume of a substance depends nonlinearly on the energy of an ionization radiation absorbed by this volume. Currently, the most studied nonlinear scintillation phenomenon is the nonproportionality of scintillation detectors during the registration of $\gamma$-radiation \cite{1,2,3,4,5,6,7,7a,8,9,10}. In this case, the light yield $LY$ of a scintillator is different for different energies of exciting single $\gamma$-quanta ($LY=E_{lum}/E_{abs}$, where $E_{lum}$ – energy of emitted luminescence, $E_{abs}$ – energy of absorbed ionizing radiation).

This phenomenon is explained by a dependence of luminescence parameters (intensity, decay time, etc) on the EEs volume density $n$~\cite{1,2,3,4,5,6,7,7a}. An increase in $n$ leads to the increase in the probability of EEs nonradiative quench due to interactions between EEs. This leads to a significant reduction in $LY$ at high $n$ ($\sim$10$^{19}$~cm$^{-3}$ and higher)~\cite{8}. Simultaneously, values of $n$ in tracks produced by $\gamma$-quanta with different energies are different. Thereby, a combined effect of the two factors – a $LY(n)$ dependence and a dependence of both $n$ values and its spatial distributions in tracks on the energy of an ionizing particle leads to the fact that $LY$ is changed for $\gamma$-quanta having different energies. 


Another important class of nonlinear scintillation phenomena is associated with a dependence of the luminescence kinetic parameters on $n$~\cite{15,16,17,18,19,20,21,22,23,24,24a}. In particular, it is found~\cite{15} that an excitation of luminescence by a synchrotron radiation with quanta energies 3-62 keV have a following feature. If a photon flux is higher than 10$^{16}$ ph.$\times$s$^{-1}\times$cm$^{-2}$, the scintillation decay times of many oxide, fluoride and alkali-halide scintillators are significantly decreased in comparison to an excitation with low photon fluxes. Similar phenomena are observed in luminescence from YAG:Ce \cite{16}, BaF$_{2}$ \cite{16,17}, CdWO$_{4}$ \cite{17} and CaWO$_{4}$ \cite{18} excited by x-ray radiation of a free electron laser. Also, it is found in Ref.~\cite{19} that the decay time of CsI:Tl scintillation excited by $\gamma$-quanta with the energy of 6 keV is shorter than the luminescence decay time excited by 661.7 keV $\gamma$-quanta.

The luminescence decay time shortening of self-trapped excitons (STE) in CdWO$_{4}$ induced by an increase in the exciting laser radiation intensity is studied in Refs.~\cite{20,21}. Similar phenomenon is observed in STE luminescence from CsI \cite{22}. The obtained results were
described quantitatively using the conception of an increase in the non-radiative quenching of excitons due to their dipole-dipole interaction with increasing $n$ \cite{20,21,22}. A theoretical description of the changes in the CsI:Tl scintillation kinetics excited by $\gamma$-quanta with different
energies is given in Ref. \cite{23}. A proposed model takes into account hot and thermalized EEs diffusion, effect of electric field on EEs evolution, EE trapping, their nonlinear quenching and radiative recombination.

In general, it should be noted that only a few nonlinear scintillation phenomena are currently well understood. In particular, mechanisms of nonlinear scintillation phenomena and their effect on the spectral and kinetic characteristics of scintillation are studied insufficiently. Thereby, their studies represent an important challenge for fundamental research.
Simultaneously, studies of nonlinear scintillation phenomena generate significant interest for practical applications because, on the one hand, they deteriorate parameters of scintillation detectors (amplitude resolution, proportionality, $LY$, etc \cite{4,14}) and thereby they should be
suppressed. On the other hand, a dependence of kinetic parameters on $n$ can be used to create detectors distinguishing a type of registered radiation \cite{24}. Indeed, particles of different types, for example, electrons and $\alpha$-particles with equal initial energies creates in a matter the ionization tracks with different spatial distribution of $n$.

Thereby, development of new scintillation materials with a high $LY$ and considerable scintillation nonlinearity generates an interest for practical applications. In this regard, rare-earth oxyorthosilicates with chemical formula Re$_{2}$SiO$_{5}$ (Re – trivalent rare-earth ion) are promising materials because they have a high chemical and radiative stability, a high mechanical strength, a bright luminescence and a fairly high mass density. Therefore, they are now widely used as high-performance scintillators (Lu$_{2}$SiO$_{5}$:Ce and its modifications\cite{27,28,29,30}, Gd$_{2}$SiO$_{5}$:Ce \cite{31,32,33}, etc), active laser media\cite{34}, quantum memory devices~\cite{34a}, etc. In this work, the spectral and kinetic properties of luminescence from new oxyorthosilicate Sc$_{1.318}$Y$_{0.655}$Si$_{1.013}$O$_{4.987}$ crystal\cite{35} excited by high-energy particles - $\gamma$-quanta with the energy 661.7 keV and electrons with the energies of 50-300 keV. The nonlinear scintillation properties of this material are observed and discussed.
 
\begin{table*}
\caption{\label{tab:table3}Kinetic parameters of intrinsic CL from (Y$_{2}$Sc$_{1}$)$_{0.(3)}$(Sc)[Si]O$_{5}$ at room temperature (CL excited by an electron beam without passing through a copper foil)}
\begin{tabular}{llllll}
Wavelength, nm &\multicolumn{4}{l}{Parameters of fitting by Eq. (1) in the region of
0-50 $\mathrm{\mu}$s} &Parameter of \\
&&&&&fitting by Eq. (2) in\\
&&&&&the region of 0-3 $\mathrm{\mu}$s\\
&$a_{1}$&$a_{2}$&$\tau_{1}, ns$&$\tau_{2}, ns$&$\tau$, ns\\
\hline
260&-&-&1050$\pm$20&1490$\pm$20&1099$\pm$6\\
270&0.56&0.44&1023$\pm$6&1522$\pm$8&1081$\pm$3\\
280&0.57&0.43&1031$\pm$2&1520$\pm$5&1090$\pm$2\\
290&0.57&0.43&1045$\pm$1&1539$\pm$3&1101$\pm$1\\
295&0.50&0.50&1027.3$\pm$0.9&1469$\pm$2&1084$\pm$2\\
300&0.56&0.44&1054.9$\pm$0.8&1536$\pm$2&1108$\pm$1\\
305&0.52&0.48&1057.8$\pm$0.8&1497$\pm$2&1108$\pm$1\\
310&0.56&0.44&1066.0$\pm$0.8&1596$\pm$2&1130$\pm$1\\
320&0.58&0.42&1087.0$\pm$0.7&1657$\pm$2&1155$\pm$1\\
330&0.59&0.41&1098.4$\pm$0.7&1727$\pm$3&1175$\pm$1\\
335&0.61&0.39&1096.2$\pm$0.7&1813$\pm$3&1187$\pm$1\\
340&0.60&0.40&1106.9$\pm$0.7&1788$\pm$3&1193$\pm$1\\
345&0.58&0.42&1105.5$\pm$0.7&1752$\pm$2&1190$\pm$1\\
350&0.58&0.42&1126.5$\pm$0.7&1790$\pm$2&1211$\pm$1\\
360&0.59&0.41&1121.4$\pm$0.7&1868$\pm$3&1222$\pm$1\\
365&0.56&0.44&1109.6$\pm$0.7&1870$\pm$3&1223$\pm$1\\
370&0.53&0.47&1157.2$\pm$0.7&1817$\pm$2&1248$\pm$1\\
380&0.55&0.45&1130.5$\pm$0.7&1910$\pm$2&1249$\pm$1\\
385&0.58&0.42&1106.7$\pm$0.7&2019$\pm$3&1250$\pm$1\\
390&0.58&0.42&1094.2$\pm$0.7&2088$\pm$4&1254$\pm$1\\
400&0.62&0.38&1085.1$\pm$0.6&2334$\pm$6&1278$\pm$1\\
410&0.60&0.40&1065.5$\pm$0.7&2403$\pm$7&1293$\pm$1\\
420&0.54&0.46&997.2$\pm$0.7&2380$\pm$7&1279$\pm$2\\
430&0.61&0.39&1063.0$\pm$0.6&2695$\pm$10&1335$\pm$2\\
440&0.61&0.39&1055.0$\pm$0.6&2810$\pm$11&1350$\pm$2\\
450&0.63&0.37&1046.1$\pm$0.5&3075$\pm$11&1371$\pm$2\\

\end{tabular}
\end{table*}

%

\section{Experimental detalis}\label{Experimental detalis}
\subsection{Sc$_{1.318}$Y$_{0.655}$Si$_{1.013}$O$_{4.987}$ crystal}

A crystal with the chemical formula Sc$_{1.318}$Y$_{0.655}$Si$_{1.013}$O$_{4.987}$ and the crystal-chemistry formula (Y$_{2}$Sc$_{1}$)$_{0.(3)}$(Sc)[Si]O$_{5}$ is a new chemical compound belonging to class of oxyorthosilicates developed as a quantum memory storage device \cite{35,36}. This compound is the congruent one for Sc$_{2}$SiO$_{5}$ -Y$_{2}$SiO$_{5}$ solid solutions because it corresponds to the minimal melting temperature for Sc$_{2-x}$Y$_{x}$SiO$_{5}$ (0 $\le x \le$ 2) compounds. Indeed, the (Y$_{2}$Sc$_{1}$)$_{0.(3)}$(Sc)[Si]O$_{5}$ melting temperature is of 2100 K whereas the melting temperature of Sc$_{2}$SiO$_{5}$ is of 2160 K, and Y$_{2}$SiO$_{5}$ melting temperature – 2200 K. The mass density $\rho_{M}$ of (Y$_{2}$Sc$_{1}$)$_{0.(3)}$(Sc)[Si]O$_{5}$ crystal is of 3.859 g/cm$^{3}$. Details of (Y$_{2}$Sc$_{1}$)$_{0.(3)}$(Sc)[Si]O$_{5}$ crystalline structure are published in Ref. \cite{36}. The crystal belongs to the C2/c space group. Sc and Y cations are distributed over the two independent crystallographic positions with the coordination numbers (CN) of 6 and 7, correspondingly. All Y$^{3+}$ ions are situated in the similar crystallographic positions and occupy 2/3 positions having CN=7. Simultaneously, Sc$^{3+}$ ions occupy all positions with CN=6 and 1/3 positions with CN~=~7. The crystal (Y$_{2}$Sc$_{1}$)$_{0.(3)}$(Sc)[Si]O$_{5}$ was grown using the Czochralski method from an iridium crucible in the 99\%Ar + 1\%O$_{2}$ atmosphere from Sc$_{2}$O$_{3}$, Y$_{2}$ O$_{3}$ and SiO$_{2}$ initial materials with the mass fractions of the primary substances at least 99.99 wt.\%. The crystal was cut into cuboids with the dimensions of 3 $\times$ 3 $\times$ 4 mm$^{3}$. Luminescence of these cuboids is studied in this work. Actual chemical composition of studied samples was measured using the technique described in Ref.~\cite{36}. Averaging of the compositions measured in different spatial points of the samples shows that the crystals have   the following chemical formula: Sc$_{1.318\pm0.011}$Y$_{0.655\pm0.005}$Si$_{1.013\pm0.003}$O$_{4.987\pm0.003}$. 

\subsection{Luminescence measurements}

All studies of (Y$_{2}$Sc$_{1}$)$_{0.(3)}$(Sc)[Si]O$_{5}$ crystal luminescence characteristics in this work were carried out in air at atmospheric pressure and room temperature.

\subsubsection{CL measurements}

An effective method of studies the scintillation processes at a high EE density including nonlinear scintillation effects is an analysis of cathodoluminescence (CL) from a substance excited by a high-power pulsed electron beam \cite{37,37a,38,38a,39,40,25,26,41,42,43}. A setup for CL excitation and a scheme of measuring the electron beam parameters by an x-ray sensor in this work are similar to that used in Ref.~\cite{25}. A CL excitation was carried out by a pulsed electron beam generated by a RADAN-EKSPERT electron accelerator equipped with an IMA3-150E vacuum accelerating diode~\cite{38,44,45}. A kinetic energies of electrons were of 50-300 keV. Every voltage pulse applied to the accelerating diode led to generation of an electron pulse with the number of 10$^{10}$ -10$^{12}$ electrons during the explosive electron emission (EEE) \cite{46}. The cross-sectional area of the beam was of 2~cm$^{2}$. Each electron pulse was composed of two components – the “fast” one and the “slow” one. The “fast” component with the duration of $\sim$50 ns was consisted of seven to eight electron bunches with a $\sim$1 ns duration and the time between the nearest bunches was of $\sim$5 ns \cite{25,26,47}. After that, the “slow” stage dominated with the duration of $\sim$1 $\mathrm{\mu}$s. The beam current in the “slow” stage was of $\sim$10$^{3}$ times fewer than in the “fast” stage \cite{25,26}. A repetition rate of the accelerator was of 1 Hz.

A temporal dependence $I_{CL}(t)$ of the CL intensity (kinetics) at different wavelengths was measured using the setup described in Refs.~\cite{25,47}. CL was directed using a 15 m quartz optical fiber into the input slit of a Solar TII MS2004 monochromator with a 1200 gr/mm diffraction grating. Behind the output slit of the monochromator, a Hamamatsu H3695-10 photomultiplier tube (PMT) was mounted. A photocurrent was measured by a Tektronix MSO6-2500 oscilloscope with a bandwidth of 2.5 GHz. A total temporal resolution of the system was of $\sim$1 ns. The long optical fiber was used in order to suppress a high-power electromagnetic noise generated by the high-current electron accelerator. A transmission region of the fiber was of 250-1000 nm.

The energy spectral density of CL (spectrum) was measured similar to Refs.~\cite{25,47} using an OCEAN FLAME-S-XR1-ES spectrometer in the region of 200-1000 nm. CL was directed into the entrance slit of this spectrometer by a 1 m quartz optical fiber. In this case, we used a short fiber because this spectrometer is based on a semiconductor photodetector. Thereby, it is less sensitive to the accelerator electromagnetic noise than the PMT. An exposure time during the CL spectra measurements was of 5 s for a beam without its passage through a copper foil and 65 s in the case of a passage of an exciting electron beam through a filter based on a copper foil (see the end of this section).

Along with the CL spectrum and kinetics, during the CL measurements, an energy of an exciting electron beam was also determined. Using this value, we then obtained the volume-averaged value of $n$ in a studied crystal. A method of electron beam energy measurement and a method of volume-averaged EE density determination is completely described in Ref.~\cite{25}. In this work, similar methods and setup were used. An energy of exciting electron beam was measured by an analysis of x-ray radiation induced by the exciting electron beam. This x-ray radiation was measured using an x-ray sensor based on a plastic polystyrene: POPOP+$\pi$- terphenyl scintillator. This scintillator was covered with a 20 mkm aluminum foil and was mounted at a distance of 3 cm from the beam axis. The aluminum foil was used as a filter passing the x-ray radiation with the energy quanta higher than 4 keV.

Thereby, the sensor registered the hard component of x-ray radiation. The luminescence from the sensor was transported through the separate channel of the 15 m fiber into the entrance window of a FEU-85 PMT. Its photocurrent was measured by the same Tektronix MSO6-2500 synchronously with the CL signal. Let us denote a temporal dependence of the x-ray intensity as $I_{REF}(t)$. The x-ray energy $E_{REF}$ is proportional to the integral of $I_{REF}(t)$ over time - $E_{REF} \propto \int_{0}^{\tau_{b}}{I_{REF}(t)dt}$, where $\tau_{b}$= 1000 ns – the time of electron beam existence. Previously \cite{25} we shown that $E_{REF}$ in the conditions of our experiment is directly proportional to energy of exciting electron beam.

Along with the CL characteristics of the crystal studied by an electron beam generated by the vacuum diode, we also studied the CL characteristics from a crystal excited by an electron beam after the passing through a barrier filter made of a copper foil with the thickness of 40~$\mathrm{\mu}$m. This filter insignificantly transforms the beam electron energy distribution function (EEDF). However, this filter significantly ($\sim$1 order of magnitude) reduces the flux of the exciting electron beam and thereby reduces $\bar{n}$ (volume-averaged density of EEs) created by the beam.

\subsubsection{RL measurements}

Radioluminescence (RL) from a (Y$_{2}$Sc$_{1}$)$_{0.(3)}$(Sc)[Si]O$_{5}$ crystal was excited by the monochromatic $\gamma$-radiation with the photon energy of 661.7 keV emitted by a $^{137}$Cs radioactive source. In this work, we measured an empirical distribution function (EDF) of the RL energies (further in this work – amplitude spectrum). Measurements of the amplitude spectrum were made using a classical scheme. A sample was mounted directly to the PMT Hamamatsu R4125Q input window via an optical coupling with an optical grease. Separate RL pulses arisen with a frequency of 10-1000 Hz. The emission was detected by the PMT. A corresponding photocurrent was amplified by a Canberra 2007B preamp, then by a spectrometric amplifier Polon 1101. A signal from this device was registered by an Analog-to-digital converter ADC Schlumberger
JCAN-21C. A statistical analysis and formation of EDF was made using a personal computer. A duration of a data accumulation was of $\sim$1 hour. As a result, a distribution of the number of scintillations per RL energy $E_{RL}$ was obtained where $E_{RL} \propto \int I_{RL}(t)dt$ (here $I_{RL}(t)$ is a temporal dependence of the RL intensity). In order to study the RL kinetics, a similar scheme with the same PMT Hamamatsu R4125Q and an optically-coupled crystal was used. However, a signal from the PMT was measured in this case similarly to Sec.2.2.1 by a Tektronix MSO6-2500 BW oscilloscope with a bandwidth of 2.5 GHz. Using this setup, $\sim$10$^{4}$ separate $I_{RL}(t)$ was measured and analyzed.

\section{Results and discussion}\label{Results and discussion}

\begin{figure}
\includegraphics{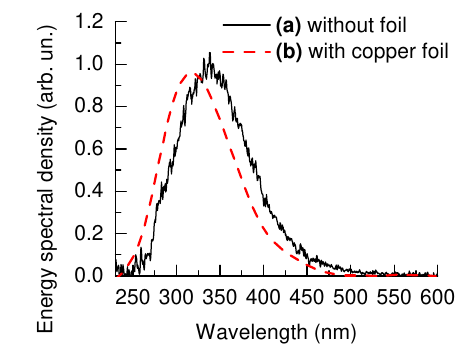}
\caption{\label{fig:epsart}
(a) CL spectrum of (Y$_{2}$Sc$_{1}$)$_{0.(3)}$(Sc)[Si]O$_{5}$ crystal (excitation by an original beam). (b) CL
spectrum of the same crystal excited by a beam passed through a 40 $\mathrm{\mu}$m copper foil.}
\end{figure}

\subsection{CL spectra of Sc$_{1.318}$Y$_{0.655}$Si$_{1.013}$O$_{4.987}$ crystal}

The CL spectrum of (Y$_{2}$Sc$_{1}$)$_{0.(3)}$(Sc)[Si]O$_{5}$ crystal is presented in Fig. 1. It is seen that at room temperature (Y$_{2}$Sc$_{1}$)$_{0.(3)}$(Sc)[Si]O$_{5}$ crystal emits a fairly bright intrinsic luminescence. The maximal energy spectral density of CL excited by the electron beam without passage through a barrier filter is at 340 nm. In general, the CL spectrum of (Y$_{2}$Sc$_{1}$)$_{0.(3)}$(Sc)[Si]O$_{5}$ is analogous to the spectra of
intrinsic luminescence from oxides containing Sc$^{3+}$ ions in their structures - LuScSiO$_{5}$ \cite{48}, Sc$_{2}$SiO$_{5}$ \cite{49,50} and Sc$_{2}$O$ _{3}$ \cite{51,52}. However, the wavelengths of the maximal luminescence energy spectral density slightly differ for these materials.

We obtained the following important result. If the fluence of the exciting electron beam is weakened by its passage through a copper foil with the thickness of 40 $\mathrm{\mu}$m, the maximum of CL spectra is shifted to 315 nm (Fig. 1b). Our Monte-Carlo calculations using GEANT4 code shown that a passing of the electron beam through the 40 $\mathrm{\mu}$m leads to the decrease in the flux of the passed beam by $\sim$20 times in comparison to the incident beam. Simultaneously, the foil leads to an insufficient little change in the EEDF of the beam which can be neglected in our analysis. A decrease in the beam flux leads to the decrease in the average EE density produced by the beam (detailed discussion in Sec. 3.5.). Thereby, Fig. 1 shows that a decrease in the EE density leads to the shift of the maximum in (Y$_{2}$Sc$_{1}$)$_{0.(3)}$(Sc)[Si]O$_{5}$ CL spectrum to shorter wavelengths.

\subsection{CL kinetics of Sc$_{1.318}$Y$_{0.655}$Si$_{1.013}$O$_{4.987}$ crystal}

\begin{figure}
\includegraphics{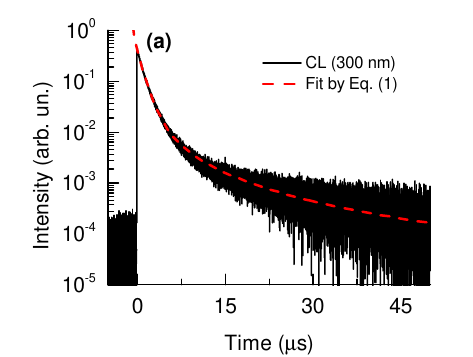}
\includegraphics{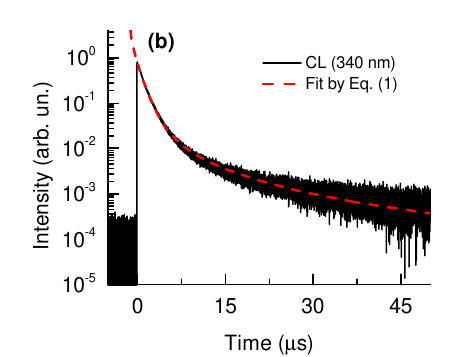}
\includegraphics{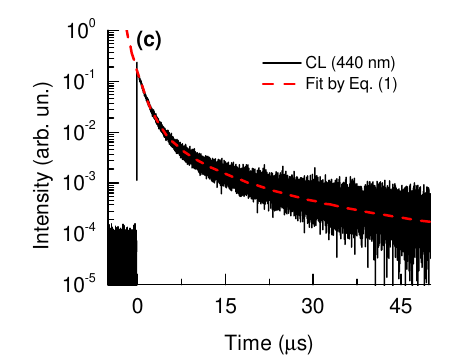}
\caption{\label{fig:epsart}
Solid curves – CL kinetics from (Y$_{2}$Sc$_{1}$)$_{0.(3)}$(Sc)[Si]O$_{5}$ crystal at 300 (a), 340 (b) and 440 (c). Dashed curves – fitting of these curves by Eq. (1).}
\end{figure}

The kinetics of CL from (Y$_{2}$Sc$_{1}$)$_{0.(3)}$(Sc)[Si]O$_{5}$ crystal at 300, 340 and 440 nm excited by an electron beam without passage through a filter are presented in Fig. 2. The CL kinetics of (Y$_{2}$Sc$_{1}$)$_{0.(3)}$(Sc)[Si]O$_{5}$ in all our experiments are composed of two stages. At the initial stage, the decay is single-exponential with the decay time of $\sim$1 $\mathrm{\mu}$s. Then the decay transforms into the hyperbolic stage. All observed curves are described fairly accurately by a following expression:

\begin{equation}
	I(t) = a_{1} \exp {\left(- \frac{t}{\tau_{1}} \right) } + \frac{a_{2}}{\left( 1 + \frac{t}{\tau_{2}} \right)^{2}},
\end{equation}

where $a_{1}$, $\tau_{1}$, $a_{2}$ and $\tau_{2}$ are the fitting parameters. 
In Fig.~2, the results of fitting for $I_{CL}(t)$ at 300, 340 and 440~nm by 
Eq.~(1) are presented. The exponential decay occurs at 0-3~$\mathrm{\mu}$s. 
The hyperbolic decay dominates from $\sim$10~$\mathrm{\mu}$s and corresponds to the Becquerel law with the power -2. This type of a radiative decay occurs due to a two-particle recombination of electrons and holes and thereby this process is a nonlinear scintillation phenomenon~\cite{1}.

The fitting parameters of $I_{CL}(t)$ by Eq. (1) are presented in Table 1. The parameters $a_{1}$ and $a_{2}$ at all wavelengths are normalized on their sum $a_{1}$ + $a_{2}$. This corresponds to the normalizing of the $I_{CL}(t)$ curves on their values at $t =$ 0. Due to this normalization, it is possible to analyze the inputs of exponential and hyperbolic decays into the $I_{CL}(t)$ dependence at different wavelengths. It is seen from Table~1, that the relation between the exponential and hyperbolic decays is nearly constant at all wavelengths. However, $\tau_{1}$ and $\tau_{2}$ have a significant spectral dependence. Indeed, the $\tau_{1}(\lambda)$ dependence ($\lambda$ is a wavelength) is a bell-shaped. The minimal value $\tau_{1}$ $\sim$1050 ns is observed at $\sim$300 and 440 nm whereas the maximal value $\tau_{1}$ = 1157.2$\pm$0.7 ns is observed at 370 nm. Simultaneously, $\tau_{2}$ increases monotonically from 1490$\pm$20 ns at 260 nm to 3075$\pm$11 ns at 450 nm.

\begin{figure}
\includegraphics{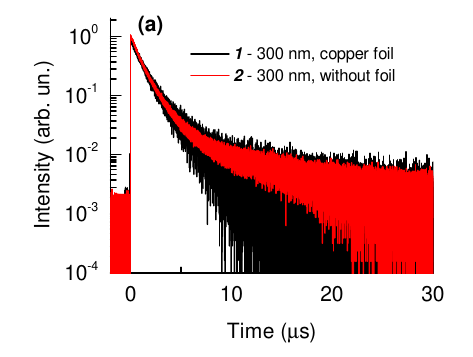}
\includegraphics{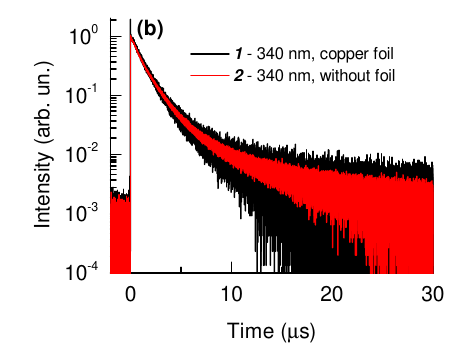}
\includegraphics{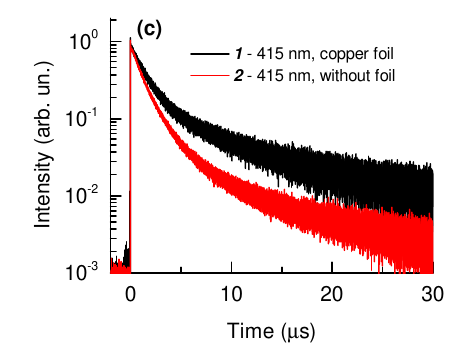}
\caption{\label{fig:epsart}
Comparison of CL kinetics from (Y$_{2}$Sc$_{1}$)$_{0.(3)}$(Sc)[Si]O$_{5}$ crystal excited by an electron beam after passed through a copper foil with the thickness of 40 $\mathrm{\mu}$m (1) and without passing through any filter (2).}
\end{figure}

As it is shown in Sec. 3.1, the spectrum CL from a (Y$_{2}$Sc$_{1}$)$_{0.(3)}$(Sc)[Si]O$_{5}$ crystal depends on the flux of incident electron beam. In order to find out if the CL kinetics likewise depends on the beam flux, the $I_{CL}(t)$ curves at 300, 340 and 415 nm measured with an electron beam without passage through a copper foil are compared with the same curves measured when the electron beam passed through the 40 mkm copper foil. This comparison is presented in Fig. 3. It is seen from Fig. 3 that at 300 and 340 nm CL kinetics are the same within the experimental error. However, the CL kinetics at 415 nm is significantly different in these cases. Indeed, a bombardment by an electron flow with smaller density occurring in the case of the passage of the
beam through the copper foil leads to longer CL decay time in comparison to the initial beam generated by the accelerator.

Thereby, the nonlinear characteristics corresponding to changes in the CL kinetics of (Y$_{2}$Sc$_{1}$)$_{0.(3)}$(Sc)[Si]O$_{5}$ crystal are observed only in the visible CL band. These properties in the ultraviolet CL are not found. The difference in dependencies of the CL kinetic parameters on $n$ for different luminescence bands shows that these bands are emitted via different mechanisms. In particular, these bands correspond to different scintillation nonlinearity mechanisms.

\subsection{RL amplitude spectrum and kinetics of Sc$_{1.318}$Y$_{0.655}$Si$_{1.013}$O$_{4.987}$ crystal}

Nonlinear scintillation properties in luminescence from (Y$_{2}$Sc$_{1}$)$_{0.(3)}$(Sc)[Si]O$_{5}$  crystal are observed in its RL as well as in its CL. The amplitude spectrum of scintillation from the studied crystal excited by $\gamma$-quanta with the energy of 661.7 keV is presented in Fig. 4. The horizontal axis in this figure shows RL energies $E_{RL}$ of each RL pulse in the relative units excited by a single $\gamma$-quantum. The vertical axis shows the number of RL pulses having the given $E_{RL}$. In Fig. 4,
at the RL energy $E_{RL}$ = 1~rel.~un., a photopeak is positioned corresponding to the internal photoelectric effect in Y atoms. During this process, a total  absorption of $\gamma$-quantum occurs. The photoelectric effect cross-section is proportional to $Z^{5}$ where $Z$ is the atomic number \cite{53}. Consequently, the photoelectric effect probability in Y atoms ($Z$ = 39) is an order of magnitude higher than that in Sc atoms ($Z$ = 21) and is several orders of magnitude higher than the photoelectric effect probability in Si ($Z$ = 14) and O ($Z$ = 8). Therefore, a photoelectric effect in these atoms can be neglected. Consequently, the photopeak is induced by the electrons ionized
from the Y$^{3+}$ K-layer with the binding energy of 17.038 keV \cite{54}. As a result of this photoelectric effect, an electron with the energy of 644.7 keV appears which excites the RL corresponding to the photopeak.

\begin{figure}
\includegraphics{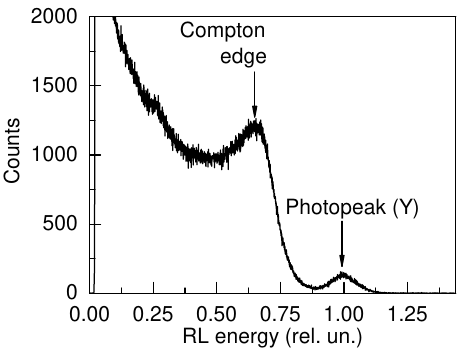}
\caption{\label{fig:epsart}
Amplitude spectrum of scintillation from (Y$_{2}$Sc$_{1}$)$_{0.(3)}$(Sc)[Si]O$_{5}$ crystal excited by the monochromatic $\gamma$-quanta with the energies of 661.7 keV.}
\end{figure}

In Fig. 4, at $E_{RL}$ = 0.65~rel.~un. there is another noticeable peak corresponding to the Compton scattering edge. The energy of Compton electrons corresponding to this edge is of 477 keV and this energy is maximal for Compton electrons induced by $\gamma$-quanta with the energies of 661.7 keV \cite{53}. A content of the hard element Y in the crystal is fairly small ($\sim$8 at.\%). Thereby, the Compton effect probability is an order of magnitude higher than the photoelectric effect probability. This fact is illustrated by the numbers of counts corresponding the the photopeak and to the Compton edge (Fig. 4).

Comparison of the photopeak positions in the amplitude spectrum excited by 661.7 keV $\gamma$-quanta corresponding to (Y$_{2}$Sc$_{1}$)$_{0.(3)}$(Sc)[Si]O$_{5}$ crystal and to CeF$_{3}$ crystal (for this material, $LY$ = 2400
photons/MeV \cite{55}) shown that the $LY$ of (Y$_{2}$Sc$_{1}$)$_{0.(3)}$(Sc)[Si]O$_{5}$ intrinsic RL is of 12000 photons/MeV. This value is fairly high and, consequently, this compound in principle can be used as a scintillation material. The photopeak shape of (Y$_{2}$Sc$_{1}$)$_{0.(3)}$(Sc)[Si]O$_{5}$ is close to the Gauss curve with the amplitude resolution of 12\%. A fairly high $LY$ of intrinsic luminescence generates an advantage of this crystal over analogues based on the activator RL due to its problems with the activator distribution non-homogeneity in grown crystals.

We found that the RL decay times excited by a photoelectron with the energy of 644.7 keV and by a Compton electron with the energy of $\approx$ 477 keV are different. In order to study the RL kinetics, the RL signal detected by a Hamamatsu R4125Q PMT was measured directly by a Tektronix MSO6-2500BW oscilloscope with the bandwidth of 2.5 GHz. In these measurements, an each $I_{RL}(t)$ curve corresponding to an each single $\gamma$-quantum was recorded separately. The duration of each oscillogram was of 100~$\mathrm{\mu}$s. A total temporal resolution in this experiment was of 3~ns. RL was excited in the same conditions as the spectrum in Fig. 4. We measured $\sim$10$^{4}$ separate $I_{RL}(t)$ curves. An amplitude spectrum calculated using this statistics, is completely similar to that shown in Fig. 4.

The obtained statistics was processed by the following way. A selection of $I_{RL}(t)$ curves corresponding to the photopeak and the Compton edge was made based on the $E_{RL}$ value of each $I_{RL}(t)$ curve. Thereby, the photopeak events fulfilled the 0.9 rel. un. $\le E_{RL} \le $ 1.1 rel. un. condition while the Compton edge events fulfilled the 0.65 rel. un. $\le E_{RL} \le $  0.9 rel. un. condition. It is seen that for an analysis of the decay kinetics of RL excited by Compton electrons, we taken into account the events from $E_{RL} = $ 0.65 rel. un. (maximum in the amplitude spectrum) to $E_{RL} $= 0.9 rel. un. where the effect of photopeak is still weak. The events with $E_{RL}<$ 0.65 rel. un. were not taken into account. Thereby, we studied the RL excited mainly by electrons with the energies near the Compton scattering edge (477 keV). Averages of these two groups of $I_{RL}(t)$ curves gives the result presented in Fig. 5. The fit of these averaged curves by an equation

\begin{equation}
	I(t) = a \exp \left( -t/\tau \right) + b,
\end{equation}

\begin{figure}
\includegraphics{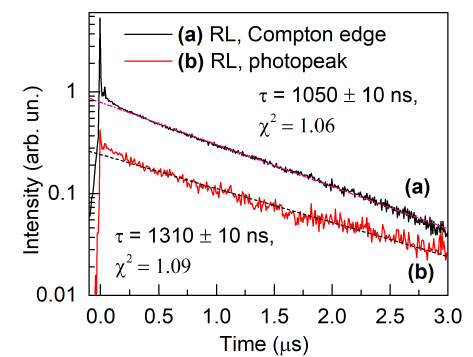}
\caption{\label{fig:epsart}
The kinetics of RL from (Y$_{2}$Sc$_{1}$)$_{0.(3)}$(Sc)[Si]O$_{5}$ crystal excited by (a) Compton electrons with the energies $\approx$ 477 keV and (b) photoelectrons with the energies of 644.7 keV. Dash and dash-dot
lines show the fits of these curves by Eq. (2).}
\end{figure}

where $a$, $b$ and $\tau$ are the fitting parameters in the time range of 0.2-3 $\mathrm{\mu}$s shows that the RL excited by a Compton electron with the energy of $\approx$ 477 keV has the decay time of $\tau =$ 1050$\pm$ 10 ns. Simultaneously, the RL excited by a photoelectron with the energy of 644.7 keV has the decay time of 1310 $\pm$ 10 ns. It is seen that an increase in the initial energy of an exciting electron leads
to the increase in the RL decay time.

\subsection{CL nonlinearity of Sc$_{1.318}$Y$_{0.655}$Si$_{1.013}$O$_{4.987}$ crystal}

According to \cite{1,2,3,4,5,6,7}, the main origin of scintillation nonproportionality and other nonlinear scintillation phenomena is on the one hand a nonlinear dependence of the electron range in a medium on the initial electron energy determining the average $n$ in a track and on the other hand a $LY(n)$ dependence governed by the EEs nonlinear quenching. A dependence of the average electron range $R$ in a medium on the initial electron energy $E_{0}$ according to Ref.~\cite{56} can be simply given by an expression $R(E_{0}) \propto E_{0}^{1.265-0.0954\ln{E_{0}}}$ ($E_{0}<$ 1 MeV, here the units of $E_{0}$ are MeVs). A track can be simply described as a cylinder~\cite{3,7a}, its length can be estimated as $R$. A track radius $r$ weakly depends on $E_{0}$ because it is determined by a free path lengths of slow secondary electrons. Their energies are of $\sim$100 eV and weakly depend on $E_{0}$~\cite{5}. For a rough estimation, we can neglect this dependence as it was made in Ref.~\cite{3}. Thereby, let us assume that the track radius is not changed with the coordinate along a track. Consequently, a volume-average EE density $n$ can be estimated by an order of magnitude as

\begin{equation}
	\bar n \approx \frac{E_{0}}{\beta E_{g} R(E_{0}) \pi r^{2}},
\end{equation}

where $E_{g}$ is the bandgap width of a material, $\beta E_{g}$ determines the threshold energy of electron–hole pair creation ($\beta \approx $ 2 $\div$ 3 is a property of a given medium). Substitution into this formula a $R(E_{0})$ dependence published in Refs.~\cite{56,56a} gives a following rough estimation of $\bar{n}$:

\begin{equation}
	\bar n \approx \frac{(Z_{x}/M_{x})}{0.412\beta E_{g} \pi r^{2} \rho_{x} (Z_{Al}/A_{Al})}E_{0}^{-0.265+0.0954 \ln E_{0}}.
\end{equation}

Here $\rho_{x}$ is the mass density of this medium in g/cm$^{2}$, $Z_{Al}$ and $M_{Al}$ are the aluminum atomic number and atomic weight, correspondingly, $Z_{x}$ and $M_{x}$ are the atomic number and atomic weight of studied material. It is seen from Eq. (4) that $\bar n$ is a decreasing function of $E_{0}$. Indeed, at $E_{0}$ = 25 keV $\bar n$ is $\sim$10 times higher than at $E_{0}$ = 1000 keV.

An increase in $n$ leads to the increase in a probability of interactions between EEs. Thereby, this process leads to an increase in EEs non-radiative quenching. As a result, it is manifested by a decrease in $LY$ and it can lead to a shortening in the luminescence decay time. In particular, an observed decrease in the RL decay time with the decrease in $E_{0}$ can be explained by an increase in nonradiative EEs quenching induced by the increase in $n$. Analogous dependence of the luminescence decay time on $n$ has the CL at 415 nm.

Simultaneously, we observed that the CL decay times at 300 and 340 nm (Fig. 3a and 3b) are not sensitive to changes in $n$ in our experiment. Thereby, nonlinear scintillation properties of (Y$_{2}$Sc$_{1}$)$_{0.(3)}$(Sc)[Si]O$_{5}$ crystal are different at different wavelengths. In order to study this
difference, we estimated the $LY(\bar n)$ at different wavelengths for this material. A method of this estimation is described in detail in \cite{25}. An experimental setup used in this work is similar to described in \cite{25} with a single exception that, in contrast to \cite{25}, in this work the samples were not covered with a diaphragm.

We applied the following method of CL nonlineartity estimation. An exciting electron beam in the accelerator is generated due to EEE \cite{46} from a surface of the vacuum diode. During EEE, the total number $N$ of electrons in each single electron shot randomly fluctuates in in our case the range from $\sim$10$^{10}$ to $\sim$10$^{12}$ particles~\cite{43,44}. Therefore, the exciting beam energy is varied by $\sim$2 orders. This leads to similar variations in $\bar n$ produced by the beam. Previously we shown \cite{25,26} that in our experiment $E_{REF}$ is proportional to the beam energy $E_{b}$.
Also, it is shown there that the dependencies of luminescence parameters on $n$ can be obtained by an accumulation of statics for different $E_{b}$. Firstly, for each electron pulse, ($E_{b}$,$E_{L}$) pairs are measured. Each of this values are calculated using the following expressions:

\begin{equation}
	E_{b} = \int_{0}^{\tilde{\tau_{b}}}{I_{REF}(t)dt}, \\
	E_{L} = \int_{0}^{\tilde{\tau}}{I_{CL}(t)dt}.
\end{equation}

We based on the relations $E_{b} \propto E_{REF}$ and $E_{L} \propto E_{CL}$. The CL yield is given by an expression $LY_{CL} = E_{L}/E_{b}$. At each wavelength, 1000  ($E_{b}$,$E_{L}$) pairs were measured. After that, the obtained data was sorted by $E_{b}$ in the ascending order. Then the data was divided into groups with 100 pairs in each group. For the each group, average values of ($E_{b}$, $E_{L}$) and $LY$ and their standard errors were calculated.

An estimation of $\bar {n}$ provided by the exciting electron beam was made by the following method. Here we give only a general scheme of calculations for brevity. In order to obtain detailed information, see Ref. \cite{25}. Regarding the Coulomb repulsion of electrons in the beam, we obtained a common expression for $\bar {n}(E_{b})$ dependence with two adjustment parameters
determined empirically from experiments \cite{25}. Simultaneously, a $LY(n)$ dependence for Bi$_{4}$Ge$_{3}$O$_{12}$ crystal was obtained using photoluminescence is published in Ref. \cite{8}. In Ref.~\cite{25} we
reported that, in our experiment, a EEE electron beam produces in media a EE distribution close to a homogeneous one. Similar distribution is created by a laser radiation. Also, the main origins of nonlinearity (dipole-dipole interaction between and Coulomb interaction between EEs, etc) are
related to EEs with the energies near the $E_{g}$~\cite{1}. Thereby, $LY(n)$ curve measured by the photoluminescence experiment and $LY_{CL}(\bar n)$ curve should be close to each other. Equating of $LY(n)$ for Bi$_{4}$Ge$_{3}$O$_{12}$ published in \cite{8} to the $LY_{CL}(\bar n)$ curve for the same crystal measured in our work leads to determination of two empirical parameters in the $\bar n(E_{b})$ dependence and therefore each $E_{b}$ value is linked to corresponding $\bar{n}$ value for Bi$_{4}$Ge$_{3}$O$_{12}$ crystal. The electron beam exciting CL in Bi$_{4}$Ge$_{3}$O$_{12}$ and in (Y$_{2}$Sc$_{1}$)$_{0.(3)}$(Sc)[Si]O$_{5}$ has the same parameters. Thereby, in order to obtain $\bar n$ for the studied crystal ($\bar n_{YSSO}$), this value can be recalculated from $\bar{n}$ value for Bi$_{4}$Ge$_{3}$O$_{12}$ ($\bar n_{BGO}$) using the following equation \cite{25}:

\begin{equation}
	\bar n_{YSSO} = \bar n_{BGO} \frac{\beta_{BGO}E_{g}^{BGO} \rho_{e}^{BGO} \ln{ 
	\left(
	1.16 \frac{\langle E_{0} \rangle + 2.9 I_{BGO}}{I_{BGO}} 
	\right) }
	}
	{\beta_{YSSO}E_{g}^{YSSO} \rho_{e}^{YSSO} \ln{ \left( 
	1.16 \frac{\langle E_{0} \rangle + 2.9 I_{YSSO}}{I_{YSSO}} 
	\right)
	}}.
\end{equation}

\begin{table*}
\caption{\label{tab:table3}Parameters of Bi$_{4}$Ge$_{3}$O$_{12}$ and (Y$_{2}$Sc$_{1}$)$_{0.(3)}$(Sc)[Si]O$_{5}$ crystals}
\begin{tabular}{lllllll}
Crystal&$Z$&$M$, Da& $\rho$, g/cm$^{3}$&$I$, eV&$E_{g}$, eV&$\beta$\\
\hline
Bi$_{4}$Ge$_{3}$O$_{12}$&529&1246&7.13&182\footnote{using the NIST ESTAR system \cite{57}} &4.2\cite{59,60}&2.38 \cite{60}\\
(Y$_{2}$Sc$_{1}$)$_{0.(3)}$(Sc)[Si]O$_{5}$ &107.67&226.59&3.859&138\footnote{Calculated using the NIST ESTAR system \cite{57}}&7.5\footnote{The value for Lu$_{2}$SiO$_{5}$\cite{58}} \cite{58} &2 \cite{49} \footnote{The value for Sc$_{2}$SiO$_{5}$ \cite{49}}
\end{tabular}
\end{table*}

In this expression, $\rho_{e} = \rho Z/M$ is the volume density of the electron number in medium, $\rho$ is its mass density, $Z$ is the atomic number and $M$ is the atomic weight. Also, in Eq.~(6), $I$ is the average ionization potential determined by the ESTAR system~\cite{57}, $\langle E_{0} \rangle$ = 120 keV is the average energy of electrons in the exciting beam. The values of parameters in Eq.~(6) used for calculation are presented in Table 2. Currently, values of $E_{g}$ and $\beta$ for (Y$_{2}$Sc$_{1}$)$_{0.(3)}$(Sc)[Si]O$_{5}$ crystal are unknown. Therefore, we use in our estimations these parameters for close materials - Sc$_{2}$SiO$_{5}$ and Lu$_{2}$SiO$_{5}$. According to Ref.~\cite{49}, $\beta E_{g} =$ 15 eV for Sc$_{2}$SiO$_{5}$. Simultaneously, for Lu$_{2}$SiO$_{5}$ $E_{g}$ = 7.5 eV \cite{58}. Thereby, if we accept that for (Y$_{2}$Sc$_{1}$)$_{0.(3)}$(Sc)[Si]O$_{5}$ $E_{g}$ = 7.5 eV and $\beta E_{g}$ = 15 eV, an electron-hole threshold is $\beta$ = 2. Also, we should recalculate $E_{b}$ from its value for Bi$_{4}$Ge$_{3}$O$_{12}$  to the value for (Y$_{2}$Sc$_{1}$)$_{0.(3)}$(Sc)[Si]O$_{5}$. Indeed, the cross-section of the bremsstrahlung radiation is proportional to $\rho Z^{2}$ for a given material \cite{53}. Consequently, $E_{b}^{YSSO} = \rho_{YSSO} Z_{YSSO}^{2} E_{b}^{BGO} /\rho_{BGO} Z_{BGO}^{2}$.

\begin{figure}
\includegraphics{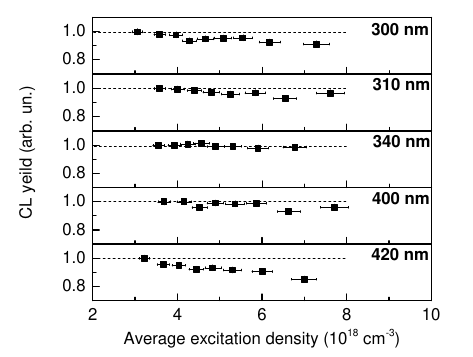}
\caption{\label{fig:epsart}
Estimations of CL yield dependence on the volume-averaged EE density for
(Y$_{2}$Sc$_{1}$)$_{0.(3)}$(Sc)[Si]O$_{5}$ crystal at different wavelengths.}
\end{figure}

The results of analysis described above are presented in Fig. 6. It should be noted that $\bar n$ values presented in Fig. 6 are the estimates and they are not the accurate values. Errors in Fig. 6 present statistical deviation but they do not take into account any systematic error which is now unknown. Therefore, $\bar n$ values in Fig. 6 are the rough estimates in order of magnitude. All curves are normalized on the  $LY_{CL}$ values at minimal $\bar n$. Simultaneously, all $LY_{CL}(\bar n)$ curves in Fig. 6 were measured in similar conditions. Positions of optical fiber, x-ray sensor and crystal were not changed. Therefore, the errors of the relations between the curves are accurately described by the statistical errors presented in Fig. 6.

It is seen from Fig. 6, that obtained $LY_{CL}(\bar n)$ curves are different at different wavelengths. In general, this situation occurs if luminescence mechanisms of different emission bands are different. For example, this case is observed for CeF$_{3}$ \cite{25}. It is concluded from Fig. 6, that in the
range of EE densities (2.5 $\div$ 8)$\cdot$10$^{18} cm^{-3}$ at 340 nm $LY_{CL}$ is nearly constant. However, at 300 and 420 nm $LY_{CL}$ falls with increasing $\bar n$. Indeed, at 300 nm $LY_{CL}$ falls by $\sim$10\% and at 420 nm it falls by $\sim$20 \%. Intermediate situations are observed at 310 and 400 nm. The fact that $LY_{CL}(\bar n)$ dependencies are different at different wavelengths is established.

\begin{figure}
\includegraphics{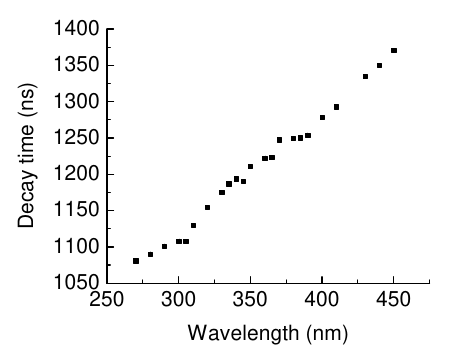}
\caption{\label{fig:epsart}
The dependence of the CL decay time on the wavelength for CL from a (Y$_{2}$Sc$_{1}$)$_{0.(3)}$(Sc)[Si]O$_{5}$ crystal. Values are obtained by fitting measured $I_{CL}(t)$ dependencies by Eq.~(2) in the time range 0$\div$3~$\mathrm{\mu}$s of the curves measured without copper foil.}
\end{figure}

\subsection{Discussion}

Let us compare the CL and RL kinetics of scintillation from (Y$_{2}$Sc$_{1}$)$_{0.(3)}$(Sc)[Si]O$_{5}$ crystal. For a correct comparison, a fitting of $I_{CL}(t)$ curves by Eq.~(2) should be made in similar conditions – in the time range 0-3 $\mathrm{\mu}$s. A $\tau_{CL}(\lambda)$ dependence calculated by this method for CL excited by the electron beam without passage through a copper foil is presented in Fig. 7. Unlike the fitting by Eq. (1), this $\tau_{CL}(\lambda)$ dependence is increased monotonically with an increase in $\lambda$. In general, it is seen from Fig. 7 that the $\tau_{CL}$ values are consistent with the RL decay times (Fig. 5).

Let us calculate $\tau$ for CL at 415~nm by the same method for excitation by the initial electron beam and the electron beam passed through the 40~$\mathrm{\mu}$m copper foil (Fig.~3c). Fitting of these dependencies by Eq.~(2) in the time range 0-3~$\mathrm{\mu}$s gives $\tau$ = 1165~$\pm$~1~ns for the initial electron beam and $\tau$ = 1377~$\pm$~3~ns in the case of the electron beam with the flux decreased by $\sim$20~times. It is seen that these decay times are close to the decay times of RL excited by 477~keV and 644.7~keV electrons. However, $\tau$ for CL at 300 and 340~nm are equal for the initial and weakened electron beam within the experimental error.

Therefore, on the one hand, a decrease in $\bar{n}$ leads to the shift of the CL spectrum to shorter wavelengths (Fig. 1). On the other hand, an increase in the energy of single particles associated with the decrease in $\bar{n}$ leads to the increase in $\tau$. However, at shorter wavelengths shorter decay
time is observed (Fig. 7). This contradiction is resolved as follows. During the RL measurements, the photosensitivity of the PMT was maximal in visible region. It should be noted that in the visible region the CL band is located whose decay time depends on the energy of exciting electrons. It means that, during the RL registration, the measured emission is mainly the luminescence from this visible band. On the other hand, the ultraviolet bands with weaker dependencies on $\bar{n}$ are less involved in the formation of the measured $I_{RL}(t)$ dependence. Moreover, a shift of the spectrum into the ultraviolet region leads to the further decrease in the input of these ultraviolet bands into the measured curve.

Despite of the fact that both the shift in CL spectrum and the change in the decay time of scintillation are induced in general by interactions between EEs at their high concentrations, we found that the mechanisms of the shift in the CL spectrum and the change in the scintillation decay time are different.  
Indeed, a shift in the spectrum with the change in $\bar {n}$ qualitatively differs from those which leads to the change in the scintillation decay time because the observed spectral shift is related to all the observed luminescence bands while the changes in the scintillation kinetics are observed only in the visible luminescence band.

Fig. 6 shows that the $LY_{CL}(\bar {n})$ dependencies are different in the region of (3 $\div$ 8) $\times$ 10$^{18}$ cm$^{-3}$ at different wavelengths. We suppose that, for $\bar{n}$ > 8 $\times$ 10$^{18}$ cm$^{-3}$, the $LY_{CL}(\bar{n})$ at different $\lambda$ are also different because an increase in $\bar{n}$ leads to the increase in the EEs interaction probability. Differences between $LY_{CL}(\bar{n})$ at different $\lambda$ are caused mainly by this interaction. Thereby, an increase in $\bar{n}$ should enhance this difference.

\begin{table*}
\caption{\label{tab:table3} Estimations of volume-averaged EE density in (Y$_{2}$Sc$_{1}$)$_{0.(3)}$(Sc)[Si]O$_{5}$ crystal in RL and CL experiments}
\begin{tabular}{ll}
Initial energy and type of the exciting electrons&Estimations of $\bar{n}$\\
\hline
$\sim$120 keV, dense flow of EEE-electrons, volume& 3.3$\times$10$^{18}$ cm$^{-3}$ (by Eq. (4))\\
distribution of EEs close to uniform one, a significant&2.5$\times$10$^{18}$ cm$^{-3}$ (by Eq. (3), $R$ calculated using~\cite{57})\\
overlapping between tracks \cite{25}&4$\times$10$^{18}$ cm$^{-3}$ (by the method \cite{25})\\
\\
$\sim$120 keV, flow of EEE-eletrons after passing the& 3.3$\times$10$^{18}$ cm$^{-3}$ (by Eq. (4))\\
40 $\mathrm{\mu}$m copper foil, individual tracks with a weak&2.5$\times$10$^{18}$ cm$^{-3}$ (by Eq. (3), $R$ calculated using~\cite{57})\\
overlapping&0.2$\times$10$^{18}$ cm$^{-3}$ (by the method \cite{25})\\
\\
477 keV, Compton electron, a single track&1.6$\times$10$^{18}$ cm$^{-3}$ (by Eq. (4))\\
&1.2$\times$10$^{18}$ cm$^{-3}$ (by Eq. (3), $R$ calculated using~\cite{57})
\\
\\
644.7 keV, photoelectron, a single track&1.4$\times$10$^{18}$ cm$^{-3}$ (by Eq. (4))\\
&1.1$\times$10$^{18}$ cm$^{-3}$ (by Eq. (3), $R$ calculated using~\cite{57})\\
\end{tabular}
\end{table*}

Estimations of $\bar{n}$ for different experiments in this work are presented in Table~3. Firstly, we used Eq.~(4) to estimate the average EE density in tracks produced by electrons with the initial energies of 120, 477 and 644.7 keV. In all calculations, we used the value of a track radius $r$ for close oxyorthosilicate crystal Gd$_{2}$SiO$_{5}$:Ce~\cite{3}. It should be noted that the $n$ values estimated using Eq. (4) are slightly overestimated because the average electron range is always slightly less than the actual track length due to the turns of tracks.
Also, we estimated $\bar{n}$ using Eq.~(3) with the range $R(E_{0})$ calculated by the NIST ESTAR system~\cite{57}. This calculation is slightly more accurate than that made by Eq.~(4). However, Eq.~(4) is an analytical one and thereby it provides better understanding of physical process. In the case of a (Y$_{2}$Sc$_{1}$)$_{0.(3)}$(Sc)[Si]O$_{5}$ crystal, the values of $R(E_{0})$ calculated by the NIST ESTAR system are slightly less than that calculated by expressions published in Refs.~\cite{56,56a}.

Secondly, we estimated $\bar{n}$ using the method proposed in Ref.~\cite{25} (see Sec. 3.4). A high-power electron beam with $N$ $\sim$10$^{10}$-10$^{12}$ produces nearly homogeneous EE distribution due to the significant role of EE diffusion \cite{25}. Indeed, the surface area of the beam is of 2 cm$^{2}$. Thereby, an estimation of the typical distance between the closest tracks $\delta r$ can be obtained by $\delta r$ $\sim$ $\sqrt{S/N} \approx$~10-100 nm. Simultaneously, a typical EE diffusion length during the time of $\sim$1~ns corresponding to the time of the electron beam existence is of $\sim$100~nm \cite{5}. Therefore, the tracks strongly overlap.

An example of the EDF of $\bar{n}$ values produced by the electron beam in this work is  presented in Fig.~8. The method of $\bar{n}$ determination is described briefly in Sec. 3.4. and in details in Ref.~\cite{25}. To obtain this distribution, 991 pulse of accelerator was analyzed. It is seen from Fig. 8 that the minimal value of $\bar {n}$ in this figure is of 2 $\times$ 10$^{18}$ cm$^{-3}$, the most probable $\bar{n}$ value is of 4 $\times$ 10$^{18}$ cm$^{-3}$ and the average value is of 4.6 $\times$ 10$^{18}$ cm$^{-3}$. Simultaneously, for a single track, $\bar{n}$~=~3.3 $\times$ 10$^{18}$ cm$^{-3}$ (estimation by Eq.~(4)) and $\bar{n}$~=~2.5 $\times$ 10$^{18}$ cm$^{-3}$ (estimation by Eq.~(3) whith $R$ calculated by the NIST ESTAR). Thereby, mainly, in the case of EEs created by the high-power electron beam, $\bar{n}$ is higher than the $\bar{n}$ in a single track. This fact shows that the tracks are significantly overlapped. Therefore, in this case, $\bar{n} \sim $ 4 $\times$ 10$^{18}$ cm$^{-3}$.

\begin{figure}
\includegraphics{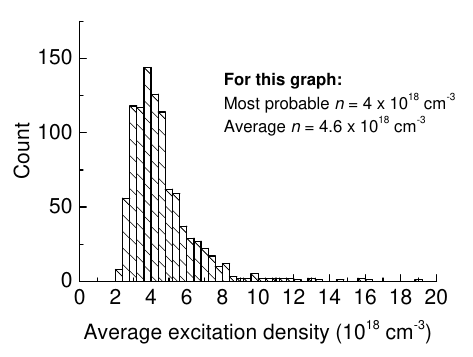}
\caption{\label{fig:epsart}
A typical view of EDF for volume-averaged EE densities created by an electron beam in (Y$_{2}$Sc$_{1}$)$_{0.(3)}$(Sc)[Si]O$_{5}$ crystal without passing through a copper foil.}
\end{figure}

Excitation of CL by the electron beam after passing through the 40 $\mathrm{\mu}$m copper foil differs from that produced by the initial electron beam. Indeed, in the case of electron beam passed through the foil, $N$ is less by $\sim$20 times. Simultaneously, the EEDF is changed very little therefore we assume that the average energy of particles in the beam is not changed ($\langle E_{0} \rangle \approx$120 keV). Consequently, for the excitation by the beam passed through the foil, $\delta r \sim$ 40-400~nm and an overlapping between the tracks still occurs but its role is significantly weaker. The most of tracks in this case are individual and they do not interact. This fact is proved by the following analysis. Division of the $\bar{n}$ value calculated for the initial electron beam by 20 gives the EE density of 0.2~$\times$10$^{18}$~cm$^{-3}$. This value is an order of magnitude smaller than the $\bar{n}$ for the individual track (3.3$\times$ 10$^{18}$ cm$^{-3}$ according to Eq.~(4) and 2.5$\times$ 10$^{18}$ cm$^{-3}$ according to Eq.~(3) with the range calculated by the NIST ESTAR). Thereby, in this case, the individual non-interacting tracks are formed with the average EE density of $\sim$3$\times$ 10$^{18}$ cm$^{-3}$ in each of them. Consequently, a correct EE density estimation for the excited area in this experiment is of $\sim$3$\times$ 10$^{18}$ cm$^{-3}$.

Therefore, in the case of CL at 415 nm, a $\sim$30\% decrease in $\bar{n}$ leads to the 18\% increase in $\tau$ from 1165~$\pm$~1~ns to 1377~$\pm$~3~ns. In the case of single $\gamma$-quanta, a $\sim$10\% decrease in $\bar{n}$ caused by the increase in the exciting electron energy from 477 keV to 644 keV leads to the 25\% increase in $\tau$ from 1050$\pm$~10~ns to 1310$\pm$~10~ns. It is seen that the results obtained from CL and RL are consistent. A difference can be caused on the one hand by the fact that the EEDF in CL is wide and consists of electrons with different energies while the RL is excited by the monochromatic particles and on the other hand the RL was measured integrally over the wavelengths while CL was measured monochromatically.

Let us estimate typical spatial scales of EEs corresponding to the emission in the visible region. An increase in $\bar{n}$ corresponds to the decrease in typical distances between EEs. The studied phenomena are observed at $\bar{n}$ $\sim$10$^{18} cm^{-3}$. This density corresponds to the typical distances between EEs of $\sim$10 nm. Therefore, at this distances, the EEs corresponding for the visible luminescence notably interact with each other while the interaction between the EEs corresponding to the ultraviolet luminescence is much weaker. In general, the mechanisms of $LY_{CL}(\bar{n})$ dependence are also related to EEs interaction. However, our results show that in the case of (Y$_{2}$Sc$_{1}$)$_{0.(3)}$(Sc)[Si]O$_{5}$ crystal these nonlinear scintillation properties occurs via different mechanisms.

As it was mentioned in Sec. 3.1, the spectral and kinetic properties of the luminescence from (Y$_{2}$Sc$_{1}$)$_{0.(3)}$(Sc)[Si]O$_{5}$ crystal are in general close to the same characteristics of LuScSiO$_{5}$ \cite{48}, Sc$_{2}$SiO$_{5}$ \cite{49,50} and Sc$_{2}$O$_{3}$ \cite{51,52}. A mechanism of Sc$_{2}$O$_{3}$ intrinsic luminescence was studied in Ref.~\cite{51}. Authors \cite{51} did not find in the reflectance spectrum any excitonic peak. Consequently, a bright intrinsic luminescence from Sc$_{2}$O$_{3}$ at 340 nm cannot currently be related to STE radiative decay. Authors \cite{51} noted that the characteristics of this luminescence are typical for a tunneling recombination of spatially correlated electrons and holes captured in traps. A similar conclusion was made in Ref.~\cite{50} in regard with intrinsic luminescence of Sc$_{2}$SiO$_{5}$.

A similarity in the spectral and kinetic parameters of luminescence from (Y$_{2}$Sc$_{1}$)$_{0.(3)}$(Sc)[Si]O$_{5}$ to the same parameters for LuScSiO$_{5}$, Sc$_{2}$SiO$_{5}$ and Sc$_{2}$O$_{3}$ shows that the mechanisms of luminescence from these materials are also similar. The recombination can occur due to transitions in crystalline defects. In particular, in nonactivated oxyorthosilicates, the emission centers can be related to oxygen vacancies \cite{61}.

\section{Conclusion}\label{Conclusion}

(Y$_{2}$Sc$_{1}$)$_{0.(3)}$(Sc)[Si]O$_{5}$ crystal has a fairly bright intrinsic luminescence at room temperature excited by both electrons with the energies of 50-300 keV and $\gamma$-quanta with the energy of 661.7 keV. The light yield of scintillation excited by the $\gamma$-quanta is of 12000 photons/MeV. Thereby, this crystal generates interest for practical applications. The Compton effect probability in the crystal is an order of magnitude higher than the photoeffect probability
due to the predomination of light elements (Sc, Si and O) in the crystal structure. This leads to formation of a pronounced Compton scattering edge in the amplitude spectrum.

In general, the spectral and kinetic parameters of intrinsic luminescence from (Y$_{2}$Sc$_{1}$)$_{0.(3)}$(Sc)[Si]O$_{5}$ are close to that for related oxides containing Sc$^{3+}$. The kinetics of (Y$_{2}$Sc$_{1}$)$_{0.(3)}$(Sc)[Si]O$_{5}$ consists of two stages. The first stage is a single-exponent one with the decay time of $\sim$1.0-1.3 $\mathrm{\mu}$s. The second stage dominates till $\sim$10 $\mathrm{\mu}$s after the beginning of the emission and corresponds to a hyperbolic Becquerel decay with the power of -2.

The scintillation from (Y$_{2}$Sc$_{1}$)$_{0.(3)}$(Sc)[Si]O$_{5}$ crystal has nonlinear properties. The decay time of RL excited by a photoelectron with the energy of 644.7~keV is of 1310~$\pm$~10~ns while in the case of a Compton electron with the energy of 477~keV the decay time is of 1050~$\pm$~10~ns. The CL decay time at 415 nm excited by the EEE beam with the average energy of 120~keV is of 1165~$\pm$~1~ns. A decrease in the fluence of the beam by $\sim$20 times leads to an increase in the CL decay time to 1377~$\pm$~3~ns. Simultaneously, the decay times at 300 and 340~nm in our experiment do not depend on the average EE density. The CL spectrum of (Y$_{2}$Sc$_{1}$)$_{0.(3)}$(Sc)[Si]O$_{5}$ crystal also depends on the average EE density. Indeed, a decrease in the flux of the exciting electron beam by $\sim$20 times leads to the shift in spectrum from 340 to 315~nm.

The obseved nonlinear parameters of intrinsic scintillation from (Y$_{2}$Sc$_{1}$)$_{0.(3)}$(Sc)[Si]O$_{5}$ are subdivided into two independent types.
A change in the temporal scintillation parameters induced by the change in the average EE density belongs to properties of the first type. These properties are found only in the visible luminescence from (Y$_{2}$Sc$_{1}$)$_{0.(3)}$(Sc)[Si]O$_{5}$ crystal. We did not found these properties in its ultraviolet luminescence.

A shift in the CL spectrum induced by the change in the average EE density belongs to properties of the second type. This effect is explained by the differences in the $LY(\bar{n})$ dependencies at different wavelengths corresponding to different emission bands including those located in the ultraviolet region. Different dependencies of $LY$ on $\bar{n}$ lead to the different energy spectral density distributions formed at different $\bar{n}$.

A decrease in the decay time of scintillation from (Y$_{2}$Sc$_{1}$)$_{0.(3)}$(Sc)[Si]O$_{5}$ crystal in the visible region accompanied by a decrease in the average EE density is explained by an increase in the nonradiative EEs quenching due to their interaction. This quenching occurs at the distances between EEs of $\sim$10 nm.

\section{Acknowledgment}
The work is supported by Russian Science Foundation (project \# 19-79-30086).

\bibliographystyle{elsarticle-num}

\bibliography{refs.bib}

\begin{thebibliography}{10}
\expandafter\ifx\csname url\endcsname\relax
  \def\url#1{\texttt{#1}}\fi
\expandafter\ifx\csname urlprefix\endcsname\relax\def\urlprefix{URL }\fi
\expandafter\ifx\csname href\endcsname\relax
  \def\href#1#2{#2} \def\path#1{#1}\fi

\bibitem{1}
A.~N. Vasil’ev, From luminescence non-linearity to scintillation
  non-proportionality, IEEE Transactions on Nuclear Science 55 (2008)
  1054--1061.
\newblock \href {https://doi.org/10.1109/TNS.2007.914367}
  {\path{doi:10.1109/TNS.2007.914367}}.

\bibitem{2}
J.~E. Jaffe, Energy and length scales in scintillator nonproportionality,
  Nuclear Instruments and Methods in Physics Research Section A: Accelerators,
  Spectrometers, Detectors and Associated Equipment. 105 (2009) 1378--1382.
\newblock \href {https://doi.org/https://doi.org/10.1016/j.nima.2007.07.059}
  {\path{doi:https://doi.org/10.1016/j.nima.2007.07.059}}.

\bibitem{3}
G.~Bizarri, W.~W. Moses, J.~Singh, A.~N. Vasil’ev, R.~T. Williams, An
  analytical model of nonproportional scintillator light yield in terms of
  recombination rates, Journal of Applied Physics 105 (2009) 044507.
\newblock \href {https://doi.org/https://doi.org/10.1063/1.3081651}
  {\path{doi:https://doi.org/10.1063/1.3081651}}.

\bibitem{4}
S.~A. Payne, W.~W. Moses, S.~Sheets, L.~Ahle, N.~J. Cherepy, N.~J. Sturm,
  S.~Dazeley, G.~Bizarri, W.-S. Choong, Nonproportionality of scintillator
  detectors: Theory and experiment, IEEE Transactions on Nuclear Science 56
  (2009) 2506--2512.
\newblock \href {https://doi.org/10.1109/TNS.2009.2023657}
  {\path{doi:10.1109/TNS.2009.2023657}}.

\bibitem{5}
G.~Bizarri, W.~W. Moses, J.~Singh, A.~N. Vasil’ev, R.~T. Williams, The role
  of different linear and non-linear channels of relaxation in scintillator
  non-proportionality, Journal of Luminescence 129 (2009) 1790--1793.
\newblock \href {https://doi.org/https://doi.org/10.1016/j.jlumin.2008.12.024}
  {\path{doi:https://doi.org/10.1016/j.jlumin.2008.12.024}}.

\bibitem{6}
Q.~Li, J.~Q. Grim, R.~T. Williams, G.~A. Bizarri, W.~W. Moses, A
  transport-based model of material trends in nonproportionality of
  scintillators, Journal of Applied Physics 109 (2011) 123716.
\newblock \href {https://doi.org/https://doi.org/10.1063/1.3600070}
  {\path{doi:https://doi.org/10.1063/1.3600070}}.

\bibitem{7}
S.~A. Payne, M.~W. W., S.~Sheets, L.~Ahle, N.~J. Cherepy, B.~Sturm, S.~Dazeley,
  G.~Bizarri, W.-S. Choong, Nonproportionality of scintillator detectors:
  Theory and experiment. ii., IEEE Transactions on Nuclear Science 58 (2011)
  3392--3402.
\newblock \href {https://doi.org/10.1109/TNS.2011.2167687}
  {\path{doi:10.1109/TNS.2011.2167687}}.

\bibitem{7a}
W.~W. Moses, G.~A. Bizarri, R.~T. Williams, S.~A. Payne, A.~N. Vasil'ev,
  J.~Singh, Q.~Li, J.~Q. Grim, W.-S. Choong, The origins of scintillator
  non-proportionality, IEEE Transactions on Nuclear Science 59 (2012)
  2038--2044.
\newblock \href {https://doi.org/10.1109/TNS.2012.2186463}
  {\path{doi:10.1109/TNS.2012.2186463}}.

\bibitem{8}
J.~Q. Grim, K.~B. Ucer, A.~Burger, P.~Bhattacharya, E.~Tupitsyn, E.~Rowe, V.~M.
  Buliga, L.~Trefilova, A.~Gektin, G.~A. Bizarri, W.~W. Moses, R.~T. Williams,
  Nonlinear quenching of densely excited states in wide-gap solids, Physical
  Review B 87 (2013) 125117.
\newblock \href {https://doi.org/https://doi.org/10.1103/PhysRevB.87.125117}
  {\path{doi:https://doi.org/10.1103/PhysRevB.87.125117}}.

\bibitem{9}
P.~R. Beck, S.~A. Payne, S.~Hunter, L.~Ahle, N.~J. Cherepy, E.~L. Swanberg,
  Nonproportionality of scintillator detectors. v. comparing the gamma and
  electron response, IEEE Transactions on Nuclear Science 62 (2015) 1429--1436.
\newblock \href {https://doi.org/10.1109/TNS.2015.2414357}
  {\path{doi:10.1109/TNS.2015.2414357}}.

\bibitem{10}
W.~Wolszczak, P.~Dorenbos, Nonproportional response of scintillators to alpha
  particle excitation, IEEE Transactions on Nuclear Science 64 (2017)
  1580--1591.
\newblock \href {https://doi.org/10.1109/TNS.2017.2699327}
  {\path{doi:10.1109/TNS.2017.2699327}}.

\bibitem{15}
V.~A. Pustovarov, A.~L. Krymov, E.~I. Zinin, Time-resolved luminescence of
  scintillation crystals under excitation by high intensity synchrotron
  radiation, Nuclear Instruments and Methods in Physics Research Section A:
  Accelerators, Spectrometers, Detectors and Associated Equipment 359 (1995)
  336--338.
\newblock \href {https://doi.org/https://doi.org/10.1016/0168-9002(94)01380-2}
  {\path{doi:https://doi.org/10.1016/0168-9002(94)01380-2}}.

\bibitem{16}
M.~Kirm, A.~Andrejczuk, J.~Krzywinski, R.~Sobierajski, Influence of excitation
  density on luminescence decay in y3al5o12:ce and baf2 crystals excited by
  free electron laser radiation in vuv, physica status solidi (c) 2 (2005)
  649--652.
\newblock \href {https://doi.org/https://doi.org/10.1002/pssc.200460255}
  {\path{doi:https://doi.org/10.1002/pssc.200460255}}.

\bibitem{17}
A.~Belsky, B.~Carré, N.~Fedorov, E.~Feldbach, J.~Gaudin, S.~Guizard,
  G.~Geoffroy, M.~De~Grazia, M.~Kirm, P.~Martin, H.~Merdji, V.~Nagirnyi,
  G.~Petite, Interaction d'impulsions vuv intenses avec les solides
  luminescents, J. Phys. IV France 138 (2006) 155--161.
\newblock \href {https://doi.org/https://doi.org/10.1051/jp4:2006138018}
  {\path{doi:https://doi.org/10.1051/jp4:2006138018}}.

\bibitem{18}
M.~Kirm, V.~Babin, E.~Feldbach, S.~Guizard, M.~De~Grazia, V.~Nagirnyi,
  A.~Vasil’ev, S.~Vielhauer, Behaviour of scintillators under xuv free
  electron laser radiation, Journal of luminescence 128 (2008) 732--734.
\newblock \href {https://doi.org/https://doi.org/10.1016/j.jlumin.2007.10.025}
  {\path{doi:https://doi.org/10.1016/j.jlumin.2007.10.025}}.

\bibitem{19}
A.~Syntfeld-Kazuch, M.~Moszynski, L.~Swiderski, W.~Klamra, N.~Antoni, Light
  pulse shape dependence on $\gamma$-ray energy in csi (tl), IEEE Transactions
  on Nuclear Science 55 (2008) 1246--1250.
\newblock \href {https://doi.org/10.1109/TNS.2008.922805}
  {\path{doi:10.1109/TNS.2008.922805}}.

\bibitem{20}
V.~Nagirnyi, S.~Dolgov, R.~Grigonis, M.~Kirm, L.~L. Nagornaya, F.~Savikhin,
  V.~Sirutkaitis, S.~Vielhauer, V.~A., Exciton–exciton interaction in cdwo4
  under resonant excitation by intense femtosecond laser pulses, Nuclear
  Instruments and Methods in Physics Research Section A: Accelerators,
  Spectrometers, Detectors and Associated Equipment 57 (2010) 1182--1186.
\newblock \href {https://doi.org/10.1109/TNS.2009.2036430}
  {\path{doi:10.1109/TNS.2009.2036430}}.

\bibitem{21}
M.~Kirm, V.~Nagirnyi, E.~Feldbach, M.~De~Grazia, B.~Carré, H.~Merdji,
  S.~Guizard, G.~Geoffroy, J.~Gaudin, N.~Fedorov, P.~Martin, A.~Vasil’ev, ,
  A.~Belsky, Exciton-exciton interactions in cdwo 4 irradiated by intense
  femtosecond vacuum ultraviolet pulses, Physical Review B 79 (2009) 233103.
\newblock \href {https://doi.org/https://doi.org/10.1103/PhysRevB.79.233103}
  {\path{doi:https://doi.org/10.1103/PhysRevB.79.233103}}.

\bibitem{22}
J.~Q. Grim, Q.~Li, K.~B. Ucer, R.~T. Williams, W.~W. Moses, Experiments on high
  excitation density, quenching, and radiative kinetics in csi:tl scintillator,
  Nuclear Instruments and Methods in Physics Research Section A: Accelerators,
  Spectrometers, Detectors and Associated Equipment 652 (2011) 284--287.
\newblock \href {https://doi.org/https://doi.org/10.1016/j.nima.2010.07.075}
  {\path{doi:https://doi.org/10.1016/j.nima.2010.07.075}}.

\bibitem{23}
X.~Lu, S.~Gridin, R.~T. Williams, M.~R. Mayhugh, A.~Gektin, A.~Syntfeld-Kazuch,
  L.~Swiderski, M.~Moszynski, Energy-dependent scintillation pulse shape and
  proportionality of decay components for csi: Tl: Modeling with transport and
  rate equations, Physical Review Applied 7 (2017) 014007.
\newblock \href
  {https://doi.org/https://doi.org/10.1103/PhysRevApplied.7.014007}
  {\path{doi:https://doi.org/10.1103/PhysRevApplied.7.014007}}.

\bibitem{24}
J.~Cang, X.~Fang, Z.~Zeng, M.~Zeng, Y.~Liu, Z.~Sun, Z.~Chen,
  Ionization-density-dependent scintillation pulse shape and mechanism of
  luminescence quenching in labr3:ce, Physical Review Applied 14 (2020) 064075.
\newblock \href
  {https://doi.org/https://doi.org/10.1103/PhysRevApplied.14.064075}
  {\path{doi:https://doi.org/10.1103/PhysRevApplied.14.064075}}.

\bibitem{24a}
K.~Wei, D.-W. Hei, J.~Liu, Q.~Xu, X.-F. Weng, X.-J. Tan, Theoretical analysis
  and experimental verification of scintillator luminescence nonlinearity based
  on carrier quenching model, Acta Physica Sinica 70 (2021) 242901.
\newblock \href {https://doi.org/10.7498/aps.70.20210820}
  {\path{doi:10.7498/aps.70.20210820}}.

\bibitem{14}
A.~Gektin, A.~N. Vasil’ev, V.~Suzdal, A.~Sobolev, Energy resolution of
  scintillators in connection with track structure, IEEE Transactions on
  Nuclear Science 67 (2020) 880--887.
\newblock \href {https://doi.org/10.1109/TNS.2020.2978236}
  {\path{doi:10.1109/TNS.2020.2978236}}.

\bibitem{27}
C.~L. Melcher, J.~S. Schweitzer, Cerium-doped lutetium oxyorthosilicate: a
  fast, efficient new scintillator, IEEE Transactions on Nuclear Science 39
  (1992) 502--505.
\newblock \href {https://doi.org/10.1109/23.159655}
  {\path{doi:10.1109/23.159655}}.

\bibitem{28}
Y.~D. Zavartsev, S.~A. Koutovoi, A.~I. Zagumennyi, Czochralski growth and
  characterisation of large ce3+lu2sio5 single crystals co-doped with mg2+ or
  ca2+ or tb3+ for scintillators, Journal of Crystal Growth 275 (2005)
  e2167--e2171.
\newblock \href
  {https://doi.org/https://doi.org/10.1016/j.jcrysgro.2004.11.290}
  {\path{doi:https://doi.org/10.1016/j.jcrysgro.2004.11.290}}.

\bibitem{29}
Y.~D. Zavartsev, M.~V. Zavertyaev, A.~I. Zagumennyi, A.~F. Zerrouk, V.~A.
  Kozlov, S.~A. Kutovoi, New radiation resistant scintillator lfs-3 for
  electromagnetic calorimeters, Bull. Lebedev Phys. Inst. 40 (2013) 34--38.
\newblock \href {https://doi.org/https://doi.org/10.3103/S1068335613020024}
  {\path{doi:https://doi.org/10.3103/S1068335613020024}}.

\bibitem{30}
V.~Kalinnikov, E.~Velicheva, A.~Rozhdestvensky, Measurement of the lyso: Ce and
  lyso:ce, ca scintillator response for the electromagnetic calorimeter of the
  comet experiment, Physics of Particles and Nuclei Letters 20 (2023)
  995--1001.
\newblock \href {https://doi.org/https://doi.org/10.1134/S1547477123050412}
  {\path{doi:https://doi.org/10.1134/S1547477123050412}}.

\bibitem{31}
K.~Takagi, T.~Fukazawa, Cerium‐activated gd2sio5 single crystal scintillator,
  Physics of Particles and Nuclei Letters 42 (1983) 43--45.
\newblock \href {https://doi.org/https://doi.org/10.1063/1.93760}
  {\path{doi:https://doi.org/10.1063/1.93760}}.

\bibitem{32}
M.~V. Belov, Y.~D. Zavartsev, M.~V. Zavertyaev, A.~I. Zagumennyi, V.~A. Kozlov,
  S.~A. Kutovoi, N.~V. Pestovskii, S.~Y. Savinov, Scintillation properties of
  oxyorthosilicate crystals gd2sio5:ce3+:ca2+, Bulletin of the Lebedev Physics
  Institute 46 (2019) 259--262.
\newblock \href {https://doi.org/https://doi.org/10.3103/S1068335619080050}
  {\path{doi:https://doi.org/10.3103/S1068335619080050}}.

\bibitem{33}
M.~V. Belov, Y.~D. Zavartsev, M.~V. Zavertyaev, A.~I. Zagumennyi, V.~A. Kozlov,
  S.~A. Kutovoi, N.~V. Pestovskii, S.~Y. Savinov, P5+ ion doped gd2sio5:ce3+
  scintillation crystal, Bulletin of the Lebedev Physics Institute 46 (2020)
  8461--8466.
\newblock \href {https://doi.org/https://doi.org/10.3103/S1068335620010029}
  {\path{doi:https://doi.org/10.3103/S1068335620010029}}.

\bibitem{34}
B.~K. Brickeen, E.~Geathers, Laser performance of yb3+ doped oxyorthosilicates
  lyso and gyso, Optics Express 17 (2009) 8461--8466.
\newblock \href {https://doi.org/https://doi.org/10.1364/OE.17.008461}
  {\path{doi:https://doi.org/10.1364/OE.17.008461}}.

\bibitem{34a}
C.~W. Thiel, T.~Böttger, R.~L. Cone, Rare-earth-doped materials for
  applications in quantum information storage and signal processing, Journal of
  Luminescence 131 (2011) 353--361.
\newblock \href {https://doi.org/10.1016/j.jlumin.2010.12.015}
  {\path{doi:10.1016/j.jlumin.2010.12.015}}.

\bibitem{35}
S.~A. Kutovoi, A.~A. Kalachev, Y.~D. Zavartsev, A.~I. Zagumennyi, V.~V.
  Voronov, V.~I. Vlasov, V.~I. Eremina, V.~F. Tarasov, Chemical composition
  based on oxyorthosilicates containing yttrium and scandium for quantum
  electronics (in russian), patent rf, num. ru2693875c1 (2019).

\bibitem{36}
L.~D. Iskhakova, A.~B. Ilyukhin, S.~A. Kutovoi, V.~I. Vlasov, Y.~D. Zavartsev,
  V.~F. Tarasov, R.~M. Eremina, The crystal structure of new quantum
  memory-storage material sc1.368y0.632sio5, Acta Crystallographica Section C:
  Structural Chemistry 75 (2019) 1202--1207.
\newblock \href {https://doi.org/https://doi.org/10.1107/S2053229619010507}
  {\path{doi:https://doi.org/10.1107/S2053229619010507}}.

\bibitem{37}
V.~I. Solomonov, Kinetics of pulsed cathodoluminescence, Optics and
  Spectroscopy 95 (2003) 248--254.
\newblock \href {https://doi.org/https://doi.org/10.1134/1.1604432}
  {\path{doi:https://doi.org/10.1134/1.1604432}}.

\bibitem{37a}
D.~I. Vaisburd, K.~E. Evdokimov, Creation of excitations and defects in
  insulating materials by high-current-density electron beams of nanosecond
  pulse duration, phys. stat. sol. (c) 2 (2005) 216--222.
\newblock \href {https://doi.org/10.1002/pssc.200460149}
  {\path{doi:10.1002/pssc.200460149}}.

\bibitem{38}
V.~I. Solomonov, S.~G. Michailov, A.~I. Lipchak, V.~V. Osipov, V.~G. Shpak,
  S.~A. Shunailov, M.~I. Yalandin, M.~R. Ulmaskulov, Clavi pulsed
  cathodoluminescence spectroscope, Laser physics 16 (2006) 126--129.
\newblock \href {https://doi.org/https://doi.org/10.1134/S1054660X06010117}
  {\path{doi:https://doi.org/10.1134/S1054660X06010117}}.

\bibitem{38a}
A.~Lushchik, C.~Lushchik, K.~Schwartz, F.~Savikhin, E.~Shablonin, A.~Shugai,
  E.~Vasil’chenko, Creation and clustering of frenkel defects at high density
  of electronic excitations in wide-gap materials, Nuclear Instruments and
  Methods in Physics Research B 277 (2012) 40--44.
\newblock \href {https://doi.org/10.1016/j.nimb.2011.12.051}
  {\path{doi:10.1016/j.nimb.2011.12.051}}.

\bibitem{39}
V.~Solomonov, A.~Spirina, \href{https://doi.org/10.5772/33833}{What is the
  Pulsed Cathodoluminescence?}, IntechOpen, 2012.
\newblock \href {https://doi.org/10.5772/33833} {\path{doi:10.5772/33833}}.
\newline\urlprefix\url{https://doi.org/10.5772/33833}

\bibitem{40}
I.~N. Ogorodnikov, M.~S. Kiseleva, V.~Y. Vostrov D.~O., Yakovlev,
  Cathodoluminescence kinetics of li6gdb3o9 crystals, Journal of Luminescence
  158 (2015) 252--259.
\newblock \href {https://doi.org/https://doi.org/10.1016/j.jlumin.2014.10.011}
  {\path{doi:https://doi.org/10.1016/j.jlumin.2014.10.011}}.

\bibitem{25}
M.~V. Belov, S.~A. Koutovoi, V.~A. Kozlov, N.~V. Pestovskii, S.~Y. Savinov,
  A.~I. Zagumennyi, Y.~D. Zavartsev, M.~V. Zavertyaev, Measurement of
  non-linearity in the cathodoluminescence yield for non-doped scintillators,
  Journal of Applied Physics 130 (2021) 233101.
\newblock \href {https://doi.org/https://doi.org/10.1063/5.0062673}
  {\path{doi:https://doi.org/10.1063/5.0062673}}.

\bibitem{26}
M.~V. Belov, S.~A. Koutovoi, V.~A. Kozlov, N.~V. Pestovskii, S.~Y. Savinov,
  A.~I. Zagumennyi, Y.~D. Zavartsev, M.~V. Zavertyaev, Interaction of
  electronic excitatoins in solid body with each other and with particles of an
  ambient gas induced during excitation of pulsed cathodoluminescence of media.
  chapter in “fast electric-explosive, electronic and 22electromagnetic
  processes in pulse electronics and optoelectronics”, ed. by g.a. mesyats,
  (in russian) (2023).

\bibitem{41}
V.~I. Solomonov, A.~V. Spirina, A.~S. Makarova, Features of the pulsed
  cathodoluminescence kinetics of neodymium ion in yttrium-aluminum garnet and
  yttria, Physics of the Solid State 64 (2022) 2088--2092.
\newblock \href {https://doi.org/10.21883/PSS.2022.13.52306.24s}
  {\path{doi:10.21883/PSS.2022.13.52306.24s}}.

\bibitem{42}
M.~I. Yalandin, V.~I. Solomonov, A.~V. Spirina, S.~A. Shunailov, K.~A.
  Sharypov, A.~S. Makarova, A.~I. Lipchak, Specific features of pulsed
  cathodoluminescence under excitation by nanosecond and subnanosecond electron
  beams, Doklady Physics 68 (2023) 50--55.
\newblock \href {https://doi.org/https://doi.org/10.1134/S1028335823020064}
  {\path{doi:https://doi.org/10.1134/S1028335823020064}}.

\bibitem{43}
V.~I. Solomonov, V.~V. Osipov, A.~S. Makarova, A.~V. Spirina, V.~V. Platonov,
  V.~A. Shitov, Luminescent response to the transformation of zinc selenide in
  ceramic synthesis, Journal of Applied Spectroscopy 91 (2024) 278–285.
\newblock \href {https://doi.org/https://doi.org/10.1007/s10812-024-01718-8}
  {\path{doi:https://doi.org/10.1007/s10812-024-01718-8}}.

\bibitem{44}
V.~N. Afanas'ev, V.~B. Bychkov, V.~D. Lartsev, V.~P. Pudov, V.~I. Solomonov,
  S.~A. Shunailov, V.~V. Generalova, A.~A. Gromov, Parameters of the electron
  beams generated by the radan-220 and radan-ekspert accelerators, Instrum.
  Exp. Tech. 58 (2005) 641–645.
\newblock \href {https://doi.org/https://doi.org/10.1007/s10786-005-0114-y}
  {\path{doi:https://doi.org/10.1007/s10786-005-0114-y}}.

\bibitem{45}
E.~H. Baksht, I.~D. Kostyrya, E.~I. Lipatov, M.~I. Lomaev, D.~V. Rybka, V.~F.
  Tarasenko, Excess-energy electrons in a nanosecond electron beam from a
  vacuum diode., Tech. Phys. 52 (2007) 489–494.
\newblock \href {https://doi.org/https://doi.org/10.1134/S1063784207040147}
  {\path{doi:https://doi.org/10.1134/S1063784207040147}}.

\bibitem{46}
G.~A. Mesyats, Explosive Electron Emission, URO-Press, Ekaterinburg, 1998,
  1998.

\bibitem{47}
M.~V. Zavertyaev, V.~A. Kozlov, N.~V. Pestovskii, A.~A. Petrov, A.~A. Rodionov,
  S.~Y. Savinov, S.~N. Tskhai, Y.~D. Zavartsev, A.~I. Zagumennyi, S.~A.
  Kutovoi, Emission of molecular nitrogen upon electron bombardment of
  pyrolytic aerogel sio2 and aluminum, JETP Letters 110 (2019) 654–658.
\newblock \href {https://doi.org/https://doi.org/10.1134/S0021364019220132}
  {\path{doi:https://doi.org/10.1134/S0021364019220132}}.

\bibitem{48}
M.~V. Belov, Y.~D. Zavartsev, M.~V. Zavertyaev, A.~I. Zagumennyi, V.~A. Kozlov,
  S.~A. Kutovoi, N.~V. Pestovskii, S.~Y. Savinov, Scintillation properties of
  new luscsio5 crystals, Bull. Lebedev Phys. Inst. 48 (2021) 97–100.
\newblock \href {https://doi.org/https://doi.org/10.3103/S1068335621040035}
  {\path{doi:https://doi.org/10.3103/S1068335621040035}}.

\bibitem{49}
V.~Y. Ivanov, V.~L. Petrov, V.~A. Pustovarov, B.~V. Shulgin, V.~V. Vorobjov,
  E.~G. Zinevich, E.~I. Zinin, Electronic excitations and energy transfer in
  a2sio5–ce (a=y, lu, gd) and sc2sio5 single crystals, Nuclear Instruments
  and Methods in Physics Research Section A: Accelerators, Spectrometers,
  Detectors and Associated Equipment 470 (2001) 358--362.
\newblock \href {https://doi.org/https://doi.org/10.1016/S0168-9002(01)01054-3}
  {\path{doi:https://doi.org/10.1016/S0168-9002(01)01054-3}}.

\bibitem{50}
V.~Y. Ivanov, E.~S. Shlygin, V.~A. Pustovarov, V.~V. Mazurenko, B.~V.
  Shul’gin, Intrinsic luminescence of rare-earth oxyorthosilicates, Phys.
  Solid State 50 (2008) 1692–1698.
\newblock \href {https://doi.org/https://doi.org/10.1134/S1063783408090217}
  {\path{doi:https://doi.org/10.1134/S1063783408090217}}.

\bibitem{51}
A.~Lushchik, M.~Kirm, C.~Lushchik, I.~Martinson, G.~Zimmerer, Luminescence of
  free and self-trapped excitons in wide-gap oxides, Journal of luminescence 87
  (2000) 232–234.
\newblock \href {https://doi.org/https://doi.org/10.1016/S0022-2313(99)00271-9}
  {\path{doi:https://doi.org/10.1016/S0022-2313(99)00271-9}}.

\bibitem{52}
O.~M. Bordun, I.~M. Bordun, Luminescence centers in sc2o3, Journal of Applied
  Spectroscopy 64 (1997) 789–792.
\newblock \href {https://doi.org/https://doi.org/10.1007/BF02678861}
  {\path{doi:https://doi.org/10.1007/BF02678861}}.

\bibitem{53}
A.~I. Akhiezer, V.~B. Berestetskii, Quantum Electrodynamics. Authorized English
  Ed., Rev. and Enl. by the Authors. Translated from the 2d Russian Ed.
  Interscience monographs and texts in physics and astronomy, v. 11
  Interscience Publishers,, Lawrence Berkeley National Laboratory, University
  of California, 1965.

\bibitem{54}
A.~C. Thompson, D.~Vaughan, X-ray data booklet, Interscience, 2001.

\bibitem{55}
A.~J. Wojtowicz, M.~Balcerzyk, E.~Berman, A.~Lempicki, Optical spectroscopy and
  scintillation mechanisms of cexla1-xf3, Physical Review B 49 (1994) 14880.
\newblock \href {https://doi.org/https://doi.org/10.1103/PhysRevB.49.14880}
  {\path{doi:https://doi.org/10.1103/PhysRevB.49.14880}}.

\bibitem{56}
L.~Katz, A.~S. Penfold, Range-energy relations for electrons and the
  determination of beta-ray end-point energies by absorption, Review of Modern
  Physics 24 (1952) 28.
\newblock \href {https://doi.org/https://doi.org/10.1103/RevModPhys.24.28}
  {\path{doi:https://doi.org/10.1103/RevModPhys.24.28}}.

\bibitem{56a}
I.~S. Grigoriev, E.~Z. Meilikhov, Physiscal values. Handbook. (Fizichseskie
  velichini, spravochnik. In Russian), Energoatomizdat, Moscow, Russia, 1991.

\bibitem{57}
Nist estar system,
  \url{https://physics.nist.gov/PhysRefData/Star/Text/ESTAR-u.html}.

\bibitem{59}
Z.~S. Macedo, A.~L. Martinez, A.~C. Hernandes, Characterization of bi4ge3o12
  single crystal by impedance spectroscopy, Materials Research 6 (2003)
  577--581.
\newblock \href
  {https://doi.org/https://doi.org/10.1590/S1516-14392003000400026}
  {\path{doi:https://doi.org/10.1590/S1516-14392003000400026}}.

\bibitem{60}
M.~Itoh, T.~Katagiri, Intrinsic luminescence from self-trapped excitons in
  bi4ge3o12 and bi12geo20: Decay kinetics and multiplication of electronic
  excitations, Journal of the Physical Society of Japan 7 (2010) 074717.
\newblock \href {https://doi.org/https://doi.org/10.1143/JPSJ.79.074717}
  {\path{doi:https://doi.org/10.1143/JPSJ.79.074717}}.

\bibitem{58}
M.~Kitaura, S.~Tanaka, M.~Itoh, Optical properties and electronic structure of
  lu2sio5 crystals doped with cerium ions: Thermally-activated energy transfer
  from host to activator, Journal of Luminescence 158 (2015) 226--230.
\newblock \href {https://doi.org/https://doi.org/10.1016/j.jlumin.2014.10.010}
  {\path{doi:https://doi.org/10.1016/j.jlumin.2014.10.010}}.

\bibitem{61}
M.~V. Belov, S.~A. Koutovoi, V.~A. Kozlov, N.~V. Pestovskii, S.~Y. Savinov,
  A.~I. Zagumennyi, Y.~D. Zavartsev, M.~V. Zavertyaev, Luminescence from oxygen
  vacancies in lu2sio5 crystal and ceramics at room temperature, Journal of
  Luminescence 263 (2023) 120155.
\newblock \href {https://doi.org/https://doi.org/10.1016/j.jlumin.2023.120155}
  {\path{doi:https://doi.org/10.1016/j.jlumin.2023.120155}}.

\end{thebibliography}

\end{document}